\documentclass[referee]{raa}            

\usepackage{graphicx,times}             
\usepackage{epsf}
\usepackage{rotating}
\usepackage{pdflscape}
\usepackage{url}
\usepackage{threeparttable}
\begin{document}

   \title{Candidates of eclipsing multiples based on extraneous eclipses on binary light curves: KIC 7622486, KIC 7668648, KIC 7670485 and KIC 8938628
}

   \volnopage{Vol.0 (200x) No.0, 000--000}      
   \setcounter{page}{1}          

   \author{Jia Zhang
      \inst{}
   \and Sheng-Bang Qian
      \inst{}
   \and Jian-Duo He
      \inst{}
   }

   \institute{Yunnan Observatories, Chinese Academy of Sciences, Kunming 650216, China; {\it zhangjia@ynao.ac.cn}\\
        \and
             Key Laboratory of the Structure and Evolution of Celestial Objects, Chinese Academy of Sciences, Kunming 650216, China\\
   }

   \date{Received~~2009 month day; accepted~~2009~~month day}

\abstract{Four candidates of eclipsing multiples, based on new extraneous eclipses found on \textsl{Kepler} binary light curves, are presented and studied. KIC 7622486 is a double eclipsing binary candidate with orbital period of 2.2799960 days and 40.246503 days. The two binary systems do not eclipse each other in the line of sight, but there is mutual gravitational influence between them which leads to the small but definite eccentricity 0.0035(0.0022) on the short 2.2799960 days period orbit. KIC 7668648 is hierarchical quadruple system candidate, with two sets of solid $203\pm5$ days period extraneous eclipses and another independent set of extraneous eclipses. A clear and credible extraneous eclipse is found on the binary light curve of KIC 7670485 which made it a triple system candidates. Two sets of extraneous eclipse of about 390 days and 220 days period are found on KIC 8938628 binary curves, which not only confirms the previous conclusion of $388.5 \pm0.3$ triple system, but also proposed a new additional objects that make KIC 8938628 a hierarchical quadruple system candidates. It is wished that the four candidates here will make some contributions to the field of eclipsing multiples.
}

   \authorrunning{J. Zhang, S. B. Qian \& J. D. He }            
   \titlerunning{Four candidates of eclipsing multiples}  

   \maketitle

%
%
\section{Introduction}           
 \label{sec:intro}

Multiple stellar systems are interesting, as well as circumbinary planets are attractive. They have variant and changing space structure, exhibiting particular physical phenomenons that binaries or single stars can not. The extraneous eclipses, what we studied in this paper, is an example. The formation and evolution of binaries depend greatly on their companion bodies around them, the close binaries need outer bodies to extract angular momentum through Kozai cycles with tidal friction (KCTF) (Kiseleva et al. 1998; Eggleton \& Kiseleva-Eggleton 2001; Fabrycky \& Tremaine 2007) to shrink to become as close as observed. So the knowledge of binaries can not be complete without their multiple environments.

More and more observations proved that binaries are often affiliated with multiples. Faherty et al. (2010) think that the ratio of triples to binaries is 3/5 and quadruples to binaries is 1/4. The statistics by Tokovinin et al. (2006) show that the fraction of spectroscopic binaries with additional companions is 96\% for period less than 3 days and 34\% for period larger than 12 days. Pribulla \& Rucinski (2006) concluded that contact binary stars exist in multiple systems with a firm lower limit of $42\%\pm5\%$. The result of D¡¯Angelo et al. (2006) is consistent with the hypothesis that all close binaries are in triple systems. Rucinski et al. (2007) contributed to the evidence that close companions is a very common for very close binaries of period less than 1 day. Not only the close binaries, the wide binaries also have companies commonly. Law et al. (2010) conclude that the fraction of multiples in wide M dwarf binaries is $45^{+18\%}_{-16\%}$.

On theoretical aspect, Fabrycky \& Tremaine (2007)'s model on KCTF supported that most or all short-period binaries have distant companions. Reipurth \& Mikkola (2012) proved that, through N-body simulations of the dynamical evolution, the widest binaries can be fairly formed from triple systems on timescales of millions of years. All these researches prompt us that, for all kinds of binaries, the companies are very common or even always there.

There are different methods to find multiple systems (Rappaport et al. 2013), the method we used is searching the extraneous eclipses, apart from those from binaries themselves, to find multiple systems. This method has been practiced successfully on \textsl{Kepler} data. KOI-126 (Carter et al. 2011) and HD 181068 (Derekas et al. 2011) are good examples. And 10 circumbinary planets (CBP) are found by this method (Welsh et al. 2015). In this paper, we present four multiple system candidates. Three of the candidates are presented with new extraneous eclipses, and the rest KIC 7622486 is analysed with new vision.

The data as the basis of the studying came from the \textsl{Kepler} mission (Borucki et al. 2010; Koch et al. 2010; Caldwell et al. 2010) which purpose is to search the exoplanet. As a byproduct, \textsl{Kepler} provided a large amount of binary light curves, that has greatly promoted the process on the binary field. Thanks to the long-term, continuous, and high-precision data, new discoveries which can not be made before were published gradually. For example, all the credible CBPs are found based on \textsl{Kepler} data. The long-term continuous changes of the depth of binary were seen clearly from \textsl{Kepler} data (e.g. KIC 7670617; KIC 8023317, Borkovits et al. (2015)). Based on \textsl{Kepler} data, Rappaport et al. (2013) and Borkovits et al. (2015) showed the clear periodic variation $O-C$ curves of several cycles length, and also provided good fittings based on new techniques. They obtained a large number of credible parameters of 65 triple systems, which reached a higher level on the period analysis of binaries, as well as on the multiple systems, both observationally and in modeling.

We search for multiple candidates through looking for the extraneous eclipses on the \textsl{Kepler} binary light curves. Different type of curves, including fragment curves, long-term curves, phase curves and curves of removing the component of binary, are examined visually for all the binary objects. The analysis method will be introduced in section 2. The candidates found are analysed one by one in alphabetical order in section 3. All the candidates are analyzed by the model of binary light curve to obtain parameters. The last section is the summarization and discussion.


\section{Data preparation}

\subsection{data acquisition}
All the light curves data in this paper are acquired from Mikulski Archive for Space Telescopes (MAST)\footnote{\url{http://archive.stsci.edu/}}, in the form of convenient compression package by \textsl{kepler} binary team. The data of October 27, 2015 version were used. However, the data of object KIC 7668648 and KIC 8938628 used here are from a older version of August 1, 2014. This is because we found that more noises were erased in the new version, it is good but unfortunately, some real eclipse signals were also erased or reduced in the meantime. For KIC 7668648 and KIC 8938628, the data of older version provided the more effective information. The Kepler Eclipsing Binary Catalog (Pr¡¦sa et al. 2011; Slawson et al. 2011; Matijevi¡¦c et al.
2012; Conroy et al. 2014) was also acquired from MAST, and this catalog provides the essential basis and guidance for our works. The catalog provide more than one period for some binary objects, which itself implies us the existence of the extraneous eclipses on the binary light curves. All the period values in this paper are from the catalog.

\subsection{data normalization}
The data used were the Pre-Search Data Conditioning (PDC) flux which purpose is `to identify and correct flux discontinuities that cannot be attributed to known spacecraft or data anomalies' (Kepler Data Processing Handbook\footnote{\url{http://archive.stsci.edu/kepler/manuals/KSCI-19081-001_Data_Processing_Handbook.pdf}}; Fraquelli et al. (2014)). Nevertheless the PDC fluxes are not the final data can be analyzed directly, the main problems are the discontinuities of quarter to quarter and the false long-term (months or longer) variations.

Hence in order to analyze the light curves of the binary (or tertiary etc.) eclipse component, all the data of each object were divided into dozens of segments, normalized for each segment and stitched together into a whole for later analysis. The segmentation point on the light curves depends on the time interval between the adjacent data points, if the time interval is larger than one day (somewhat arbitrarily), the light curves will be split at that gap. Therefore the data from different quarters were split naturally. After the segmentation, the light curve of each segment was fitted by a cubic polynomial model, and the data were divided by the fitted values to get the normalized, or flattened data. During the polynomial fitting, the jump points (e.g. the deep eclipse points, the flare points and the high dispersion points) were eliminated to obtain a better fitting. The points with fitting residuals larger than three times the standard deviations of fitting were removed, and then light curves were refitted. If the high dispersion points still exist, they will be removed before another refitting until all the high dispersion points are removed (In practice the iteration number is limited to 10 times).

This normalization process eliminated the jumps of segment to segment (including quarter to quarter), and also eliminated a part of the long term variations. These variations may be caused by the instruments, and also may be the real variations. The elimination of the long term variation will not damage, but on the contrary be conducive to, the analysis of the short term eclipse light curves.

\subsection{phase data calculation} \label{sec:phase_cal}
The phase light curves are needed for the objects in the analysis. The calculation of the phase data is as follows: 1. For the period of very small changes, a single reference epoch and period are used to calculate the phase data, and then the phase data are folded into the range of 0 to 1. For the period of large changes, the single reference epoch will lead to the misalignment of different eclipses, or lead to the inaccuracy phase value of the secondary minimums (because its adjacent primary minimum is not strictly at the integer phase). In order to cope with this situation, the minimum time of each eclipse is used as the reference epoch for its own to calculate the phase values (of course the minimum time of each eclipse is needed to be measured previously), so the phase curves in high degree of coincidence of eclipses are obtained. 2. The range of 0 to 1 is equally divided into small segments (e.g. 200 segments per unit phase). For each segment, the folded phase data obtained from step 1 are averaged to a single value (the points with large deviations are eliminated before the average). Then all the average values constitute the merged phase curve. The phase curve in this paper, if no special description, are always this kind of binning phase curve. 3. In many cases, the eclipse is very sharp compared to other parts. In order to display its profile clearly, the segments at the eclipses are need to be more denser than other parts (e.g. 10000 segments per unit phase). In the analysis of the binary modeling, this nonuniform phase data is also employed because the light curves on the eclipse part is much more important in the modeling to deserve much more weight.

\subsection{binary light curves analysis}
The binary phase light curves of the objects below were analyzed using the Wilson-Devinney programme (Wilson \& Devinney 1971; Wilson 1979, 1990; Van Hamme \& Wilson 2007; Wilson 2008; Wilson et al. 2010; Wilson 2012). All the temperatures of the primary stars are taken from the A catalogue of temperatures for Kepler eclipsing binary stars by Armstrong et al. (2014). In the analysis, the exponents in the bolometric gravity brightening law $g$ and the bolometric albedos $A$ for reflection heating and re-radiation on stars depend on the temperatures, that $g=1$ and $A=1$ for temperature higher than 7200 K, and $g=0.32$ and $A=0.5$ for lower temperatures. For here the 7200 K is designated to the boundary of the radiative and convective envelopes. The logarithmic law for limb darkening is used, and the coefficients are internally computed in the programme as a function of $T_{eff}$, $\lg g$ and $[M/H]$ based on the van Hamme (1993) table, where $T_{eff}$ is from Armstrong et al. (2014) or internal iterative calculation, $\lg g$ is calculated by the orbital period and semi-major axis that is estimated on main sequence hypothesis, and $[M/H]$ is set to be -0.2 for all objects. While the $\lg g$ and $[M/H]$ is very weak in the effect on limb darkening coefficients, so they can be with big errors.

For \textit{kepler} long-cadence data, they have long integration time of about 30 minutes for each point, so this will lead to the light curve smearing especially on sharp eclipse parts. This problem is dealt with by way of Integration by Gaussian quadrature (Wilson 2012) in the Wilson-Devinney programme, and the number of Gaussian abscissas in the running of code is set to 3. While for the short-cadence data, the smearing effects are not considered since the integration time is only about 1 minute that is short enough for our targets.

\section{Four \textsl{Kepler} binaries with extraneous eclipses}

\subsection{KIC 7622486}
Pr{\v s}a et al. (2011) and Slawson et al. (2011) give the earliest identification and light curve analysis on the object KIC 7622486, classifying it to a detached binary system. However, Borucki et al. (2011) analyzed it as a very probable false positive planetary candidates with two planets. Walkowicz \& Basri (2013) and Burke et al. (2014) obtained its properties and parameters based on the planetary candidates model premise. Rowe et al. (2014) flagged the object false-positive or false alarm in their paper Validation of Kepler's Multiple Planet Candidates. III.

\begin{figure}
\includegraphics[scale=.50]{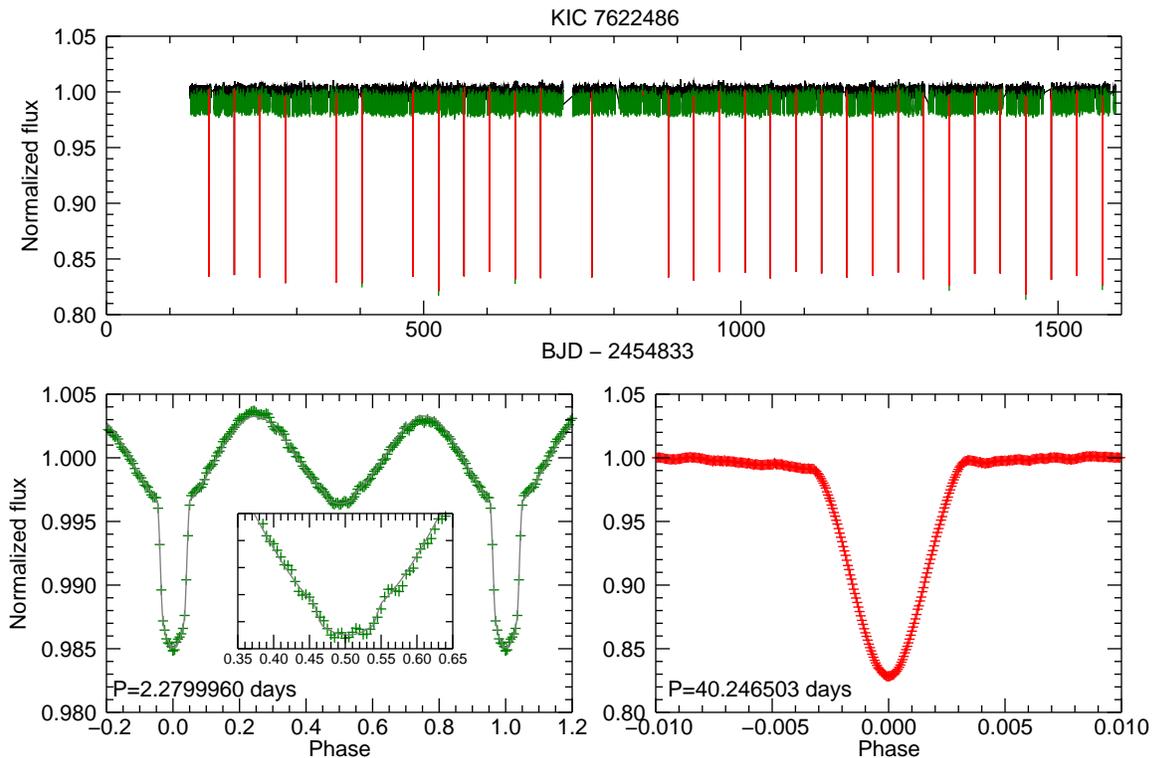}
\caption{The light curves of KIC 7622486. Upper: The whole light curves. The green and red lines are 2.2799960 days period and 40.246503 days period eclipses respectively. Lower left: The folded phase curve of 2.2799960 days period with gray fitting line. The part around phase 0.5 is magnified shown in the internal panel. Lower right: The minimum part of folded phase curve of 40.246503 days period.} \label{KIC007622486.eps}
\end{figure}

According to the shape of the light curves, the object KIC 7622486 is suggested a quadruple system consisting of two pairs of stellar binaries. This object clearly has two sets of eclipses, with strict period of 2.2799960 and 40.246503 days, see Figure \ref{KIC007622486.eps}. The folded phase curve of 2.2799960 days period is shown in the lower left panel, and it can be fitted well by a binary model of mass ratio of 0.269(0.005), see Table \ref{tab:binary_solution}. There are twists on the two sides of the secondary eclipse (about 0.45 and 0.55 in phase), both on the observational data and the theoretical light curve, which can be seen more clearly in the small internal panel within the lower left panel. The two twists are actually the boundaries of the secondary eclipses. The agreement of this little detail strengthen the reliability of the stellar binarity result of the 2.2799960 days period. The massive component star, with much bigger radius compared to the less massive one, is heavily deviated from sphere due to the mutual gravity between the two components (see the radius of star 1 in different directions in Table \ref{tab:binary_solution}). Thus, the massive star generate great light curve variations out of the eclipses.

\begin{landscape}
\begin{table}
\caption{The Binary Light Curve Solutions \label{tab:binary_solution}}

\begin{center}

 \begin{tabular}{lllllll}
  \hline\noalign{\smallskip}
 Parameters & KIC 7622486 & KIC 7668648 & KIC 7668648 & KIC 7670485 & KIC 8938628 \\
  \hline\noalign{\smallskip}
mode                                                               &  detached binary       &  detached binary     &  detached binary    &  detached binary           &  detached binary              \\
orbital eccentricity $e$                                           &  0.0035(0.0022)        &  0.0449(0.0002)      &  0.0449(0.0002)     &  0.0241(0.0007)            &  0.012(0.005)                 \\
argument of periasrtron $\omega[radians]$                          &  5.2(0.4)              &  5.445(0.004)        &  5.444(0.004)       &  1.98(0.01)                &  4.5(0.1)                     \\
orbital inclination $i[^{\circ}]$                                  &  85.68(0.07)           &  89.993(0.004)       &  89.992(0.004)      &  89.82(0.01)               &  87.21(0.04)                  \\
mass ratio $m_{2}/m_{1}$                                           &  0.269(0.005)          &  0.510(0.004)        &  0.979(0.005)       &  0.84?                     &  $0.91^a$        \\
primary temperature $T_{1}^a$                                      &  7548                  &  6449                &  6449               &  5814                      &  6412                         \\
temperature ratio $T_{2}/T_{1}$                                    &  0.49(0.02)            &  0.9659(0.0003)      &  0.9667(0.0003)     &  0.9747(0.0005)            &  0.881(0.004)                 \\
Luminosity ratio $L_{1}/(L_{1}+L_{2})$ in band kepler              &  0.999304(0.000008)    &  0.5819(0.0002)      &  0.5822(0.0002)     &  0.515(0.001)              &  0.813(0.004)                 \\
Luminosity ratio $L_{2}/(L_{1}+L_{2})$ in band kepler              &  0.000696(0.000008)    &  0.4182(0.0002)      &  0.4178(0.0002)     &  0.485(0.001)              &  0.187(0.004)                 \\
Luminosity ratio $L_{3}/(L_{1}+L_{2}+L_{3})$ in band kepler        &  0.652(0.003)          &                      &                     &  0.8275(0.0004)            &  0.33(0.01)                   \\
Modified dimensionless surface potential of star 1 $\Omega_{1}$:   &  4.04(0.02)            &  46.21(0.05)         & 46.68(0.05)         &  28.7(0.1)                 &  21.2(0.2)                    \\
Modified dimensionless surface potential of star 2 $\Omega_{1}$:   &  7.4(0.1)              &  27.0(0.2)           & 50.3(0.3)           &  24.0(0.3)                 &  30.7(0.4)                    \\
Radius of star 1 (relative to semimajor axis) in pole direction    &  0.264(0.001)          &  0.02189(0.00003)    & 0.02190(0.00002)    &  0.0360(0.0001)            &  0.0493(0.0005)               \\
Radius of star 2 (relative to semimajor axis) in pole direction    &  0.044(0.001)          &  0.0199(0.0002)      & 0.0199(0.0001)      &  0.0368(0.0005)            &  0.0307(0.0004)               \\
Radius of star 1 (relative to semimajor axis) in point direction   &  0.270(0.001)          &  0.02189(0.00003)    & 0.02190(0.00002)    &  0.0360(0.0001)            &  0.0493(0.0005)               \\
Radius of star 2 (relative to semimajor axis) in point direction   &  0.044(0.001)          &  0.0199(0.0002)      & 0.0199(0.0001)      &  0.0368(0.0005)            &  0.0307(0.0004)               \\
Radius of star 1 (relative to semimajor axis) in side direction    &  0.268(0.001)          &  0.02189(0.00003)    & 0.02190(0.00002)    &  0.0360(0.0001)            &  0.0493(0.0005)               \\
Radius of star 2 (relative to semimajor axis) in side direction    &  0.044(0.001)          &  0.0199(0.0002)      & 0.0199(0.0001)      &  0.0368(0.0005)            &  0.0307(0.0004)               \\
Radius of star 1 (relative to semimajor axis) in back direction    &  0.269(0.001)          &  0.02189(0.00003)    & 0.02190(0.00002)    &  0.0360(0.0001)            &  0.0493(0.0005)               \\
Radius of star 2 (relative to semimajor axis) in back direction    &  0.044(0.001)          &  0.0199(0.0002)      & 0.0199(0.0001)      &  0.0368(0.0005)            &  0.0307(0.0004)               \\
  \noalign{\smallskip}\hline
\end{tabular}
\tablecomments{0.86\textwidth}{a: Come from Armstrong et al.(2014). ?: means large uncertainty}
\end{center}
\end{table}
\end{landscape}

The second set of eclipses with period of 40.246503 days is not considered the third star revolving the 2.2799960 days binary. The reason is that if the eclipses of 40.246503 days is due to the occultation between the third star and the binary, then the shape of eclipses should change over time, and the times of eclipses in one period should not always be one, but should be several times often, just like the case of KIC 7668648 and  KIC 8938628 will be shown after. Since the long 40.246503 days period eclipses do not change with the phase of short period binary, it is thought that the eclipses of 40.246503 days period indicate another binary system. Moreover, the great depth, 0.15 - 0.18 in normalized flux, of the 40.246503 days period eclipses, indicates a stellar binary system rather than a system of a stellar with a planet. The extra luminosity proportion of 0.652(0.003) means the outer binary is more luminous, and more massive guessed, than the inner binary. This phase light curves of 40.246503 days are not analyzed, because the lack of secondary minimum will make the results unreliable.

Is the 40.246503 days period binary gravitational bound to the 2.2799960 days binary, or just the a blended source? Considering a small eccentricity 0.0035(0.0022) is needed to to fit the inner binary light curve, and the 2.2799960 days binary is too small to generate a eccentric orbit itself\footnote{If the 2.2799960 days binary is very young in age, it is possible that there is residual eccentricity from the initial formation, for its age is not long enough to orbit circularization. Since KIC 7622486 is a field star and do not belong to a young star cluster, it is suggested that its age is old enough to circularize its own orbit.}, this small eccentricity should be caused by the perturbation of an outer companion. So the 40.246503 days period binary is likely to be the outer companion, and that is to say, it is more suggested that KIC 7622486 is consisted of two binaries gravitational bound to each other, rather than two blended binaries in the sky.

\subsection{KIC 7668648}

Pr{\v s}a et al. (2011) and Slawson et al. (2011) made the first discovery and analysis for this object KIC 7668648. Rappaport et al. (2013) and Borkovits et al. (2015) measured and analyzed its eclipse timing variations of the 27.818590 days period, and also found some extra shallow eclipses, not due to the binary, matching the 204.8 days period $O-C$ curve which can conclude that KIC 7668648 is a triply eclipsing system. This object also has other interesting features: The object has the lowest period ratio ($P_{2}/P_{1} < 7.3$) found so far; The mutual inclination of the binary orbital plane with respect to the orbital plane of the triple is $40.9^{\circ}$, that slightly larger than the critical Kozai oscillations inclination $39.2^{\circ}$; And the eclipses of 27.818590 days period is deepening continuously and rapidly.

\begin{figure}
\includegraphics[scale=.50]{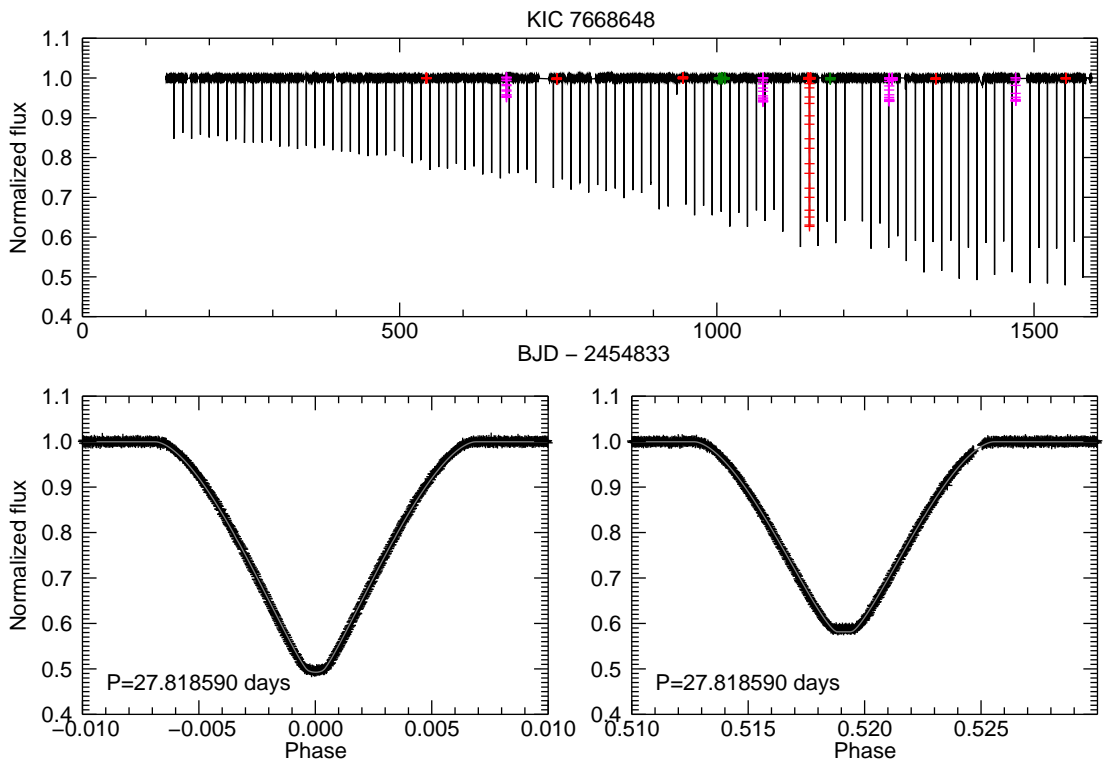}
\caption{The light curves of KIC 7668648. Upper panel: the whole long cadence light curve with extraneous eclipses in red, magenta and green plus. Points of each color stand for different sets. The red and magenta points have the same time intervals of 204.8 days. The deepest extra eclipse around 1146 days overlap a binary eclipse which can be seen clearly in the third left panel from the top. Lower panels: The 27.818590 days period binary phase curves at primary and secondary eclipses, which are calculated from the \textit{Kepler} short cadence data in 1559 to 1591 days, the gray lines are the fittings.} \label{KIC007668648.eps}
\end{figure}

\begin{figure}
\plottwo{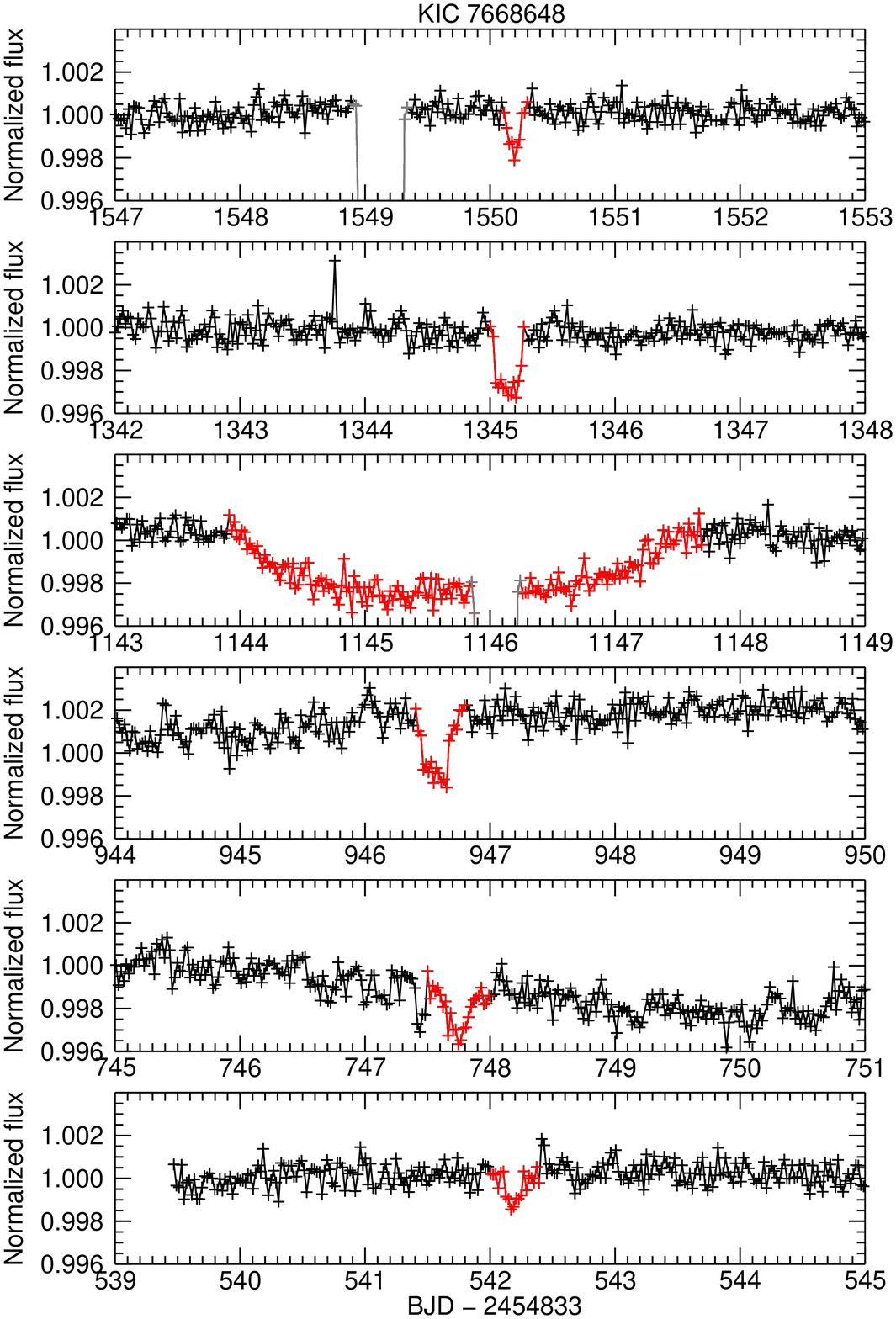}{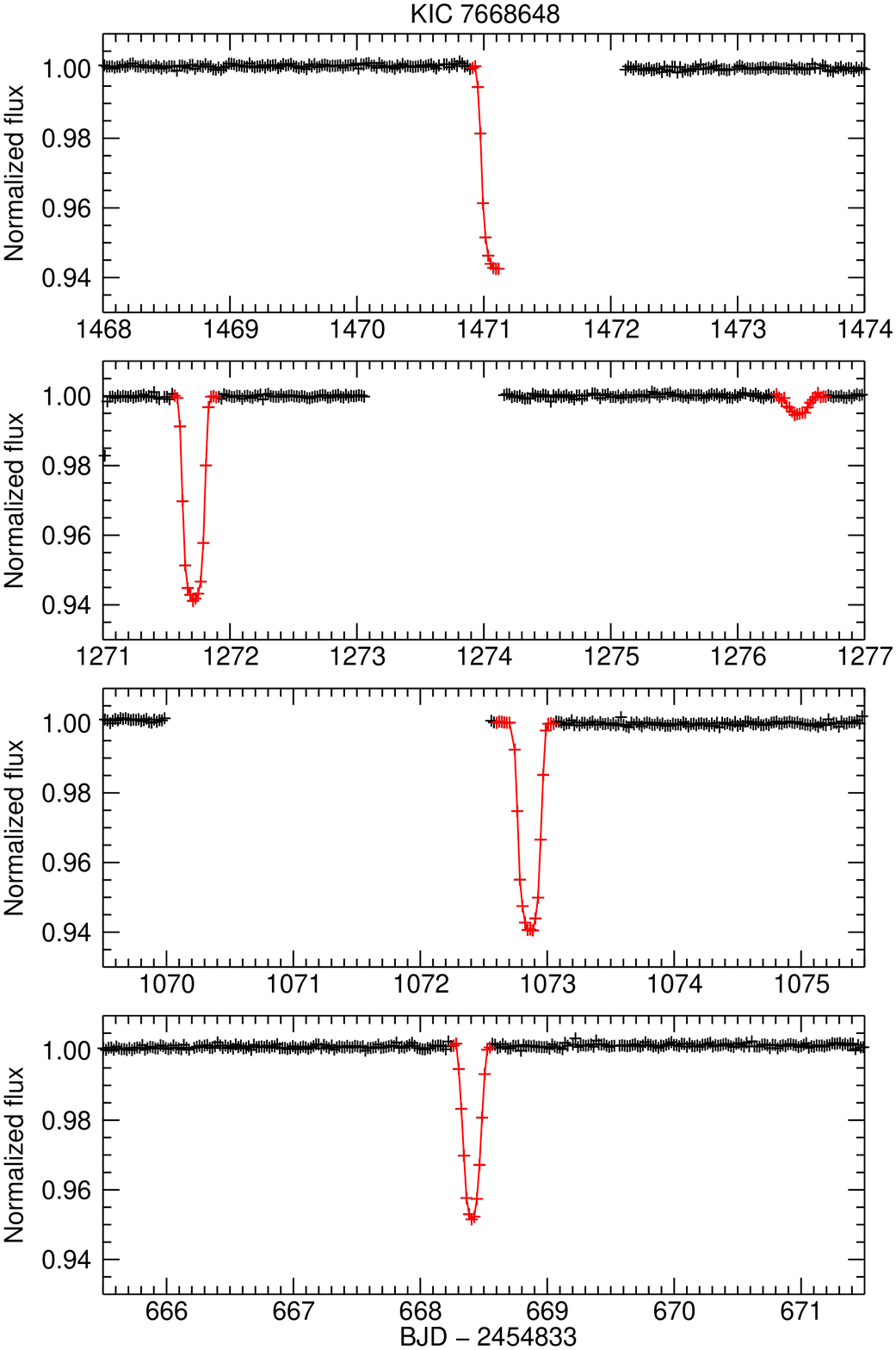}
\caption{The extraneous eclipses of KIC 7668648. The red points are the extraneous eclipses. Left panels: One set of the extraneous eclipses light curves, all these extraneous eclipses have adjacent intervals about $203\pm5$ days. The gray points, only in the first and third panel from top, indicates the binary eclipses of 27.818590 days period which are too deep and can be only shown partly. Right panels: Another set of the extraneous eclipses light curves, and the adjacent intervals are also (or times of) $203\pm5$ days.} \label{KIC007668648_extra_eclipses_1_and_2.eps}
\end{figure}

In this paper, three sets of extraneous eclipses besides the binary eclipses of 27.818590 days period are presented. Two of the extraneous eclipses, exhibiting time intervals of $203\pm5$ days, coincide with the 204.8 days period from Borkovits et al. (2015), see Figure \ref{KIC007668648.eps} and \ref{KIC007668648_extra_eclipses_1_and_2.eps}.\footnote{The data on the Figure are downloaded from August 1, 2014, instead of the updated version of April 22, 2015 or October 27, 2015. Because in the new data, the eclipse at 748 days disappears, and the eclipse of 542 days is weakened. However, according to the outer period found by Borkovits et al. (2015) and the positions of other extraneous eclipses, the eclipse on 748 days is reasonable. So it is thought that the newer data are flawed in this point. Indeed a lot of noises have been smoothed out, but the real signals are also weakened or even eliminated.} The left panels of Figure \ref{KIC007668648_extra_eclipses_1_and_2.eps} are considered to be the binary eclipsing the tertiary star, i.e. the binary blocks the light of tertiary star, because in the third panel from top where the binary eclipse superposing on the long extra eclipse. Thus the right panels are considered to the tertiary star eclipsing the binary, i.e. the tertiary star blocks the light of binary. The eclipse depth in the right panels are significantly greater than that of left panels, about 30 times difference in depth of normalized flux, which indicates that the tertiary star is much fainter than the components of binary, or the light contribution of the third body to the whole system is tiny.

\begin{figure}
\includegraphics[scale=.50]{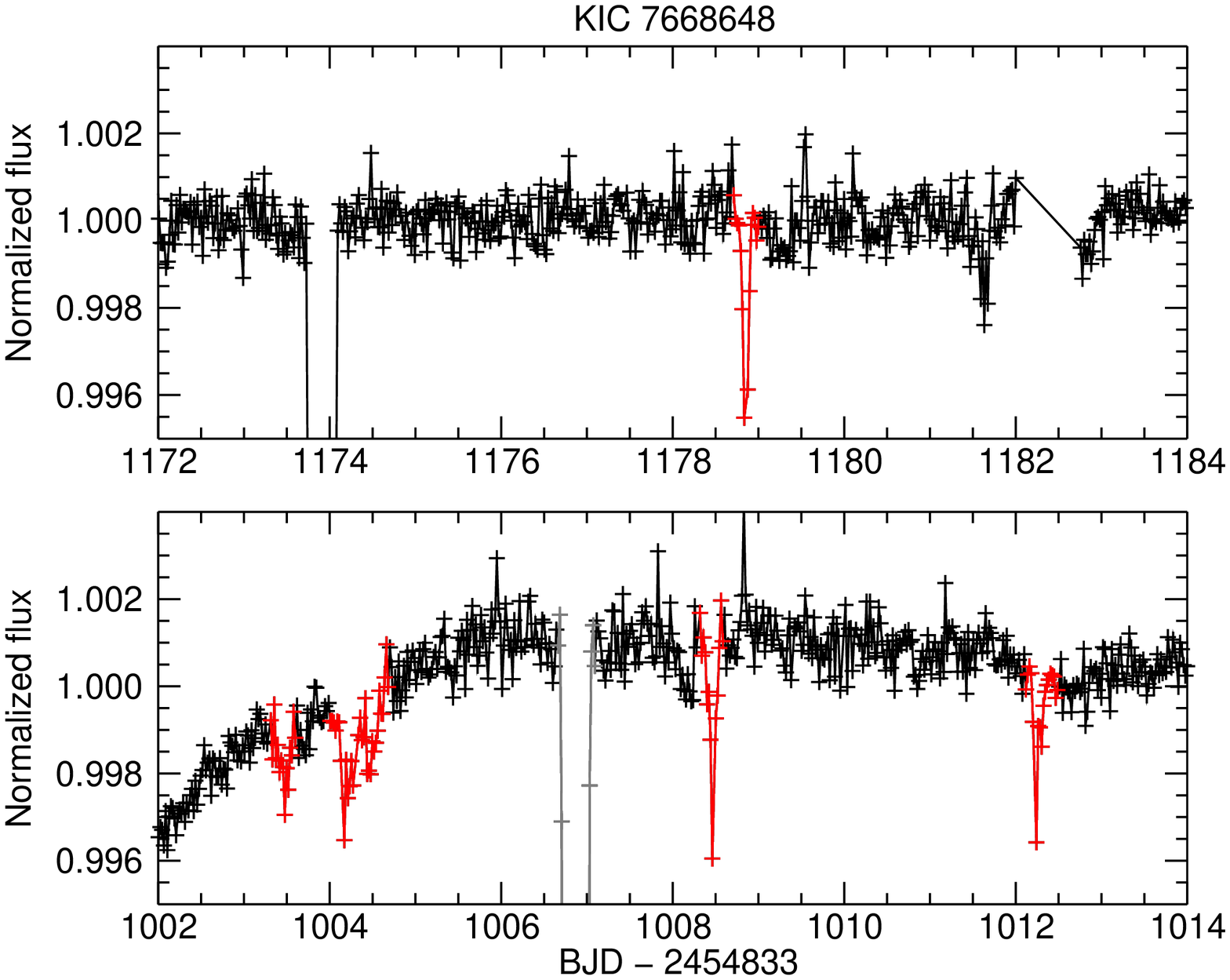}
\caption{The extraneous eclipses of KIC 7668648. The symbols are the same as Figure \ref{KIC007668648_extra_eclipses_1_and_2.eps}.} \label{KIC007668648_extra_eclipses_3.eps}
\end{figure}

The third new set of extraneous eclipses are displayed in Figure \ref{KIC007668648_extra_eclipses_3.eps}, the time interval of the two panels is about 170 days. The lower panel has more than one eclipse near a binary eclipse indicating that a fourth body revolving the 27.818590 days period binary may exist. In fact, many suspect extraneous eclipses are found but not shown here due to the large dispersions on the light curves. It is guessed that there may be more additional bodies of low mass stars, and possibly including planets.

The 27.818590 days period inner binary was analyzed with the binary model of Wilson-Devinney programme (Wilson \& Devinney 1971; Wilson 1979, 1990; Van Hamme \& Wilson 2007; Wilson 2008; Wilson et al. 2010; Wilson 2012), based on the short cadence data fortunately covering a full orbital period, and the fitting results are shown in the lower panels of Figure \ref{KIC007668648.eps} and listed in Table \ref{tab:binary_solution}. Two solutions are listed in Table \ref{tab:binary_solution}, which are only different in mass ratio and potential of star 2. The solution with mass ratio of 0.979($\pm$0.005) is suggested, because it is consistent with other parameters, i.e. the temperature ratio, the luminosity ratio and the radius ratio, which are all close to 1 between the two components. The two fitting curves corresponding to the two solutions are almost overlap so can not be distinguished in Figure \ref{KIC007668648.eps}.

The object KIC 7668648 belong to a much compact triple system, with period ratio of outer companion to inner binary is $P_{outer\:companion}/P_{inner\:binary} < 7.3$ which is the most lowest ratio so far. Thus the inner binary of KIC 7668648 is thought to has been affected by its outer companion, and then leads to the deviations from the keplerian orbit. This will arouse the doubt about applicability of binary model used on KIC 7668648. The parameters differences between Table \ref{tab:binary_solution} and those from Borkovits et al. (2015) are partly due to these deviations. Thus the parameters of Borkovits et al. (2015) is more reliable if considering this point, since they are only based on dynamical motions and without regarding to the distorted light curves.

However, there are two parameters indicating that the gravitational affection should be very limited: First, the very small but definite eccentricity $e = 0.0449(0.0002)$ in the long 27.818590 days period orbit (can be seem more intuitively from the lower right panel of Figure \ref{KIC007668648.eps} in which the phase of secondary minimum is clearly different from 0.5). Second, the small mass proportion of its outer companion, which is only 0.144(60), and the small changes of period of the inner binary (Borkovits et al. 2015). Furthermore, the photometric solution presented here is self consistent, its mass ratio, temperature ratio and radius ratio are all close to 1. In addition, the analysis program can not be converged with a positive third light value, namely the third light tends to be negative to converge, and this should be interpreted that the proportion of third light is extremely small, which is consistent to the extraneous eclipses analysis above. Therefore finally, the solutions of KIC 7668648 by the binary model are deemed to be correct in general.

\subsection{KIC 7670485}

\begin{figure}
\includegraphics[scale=.50]{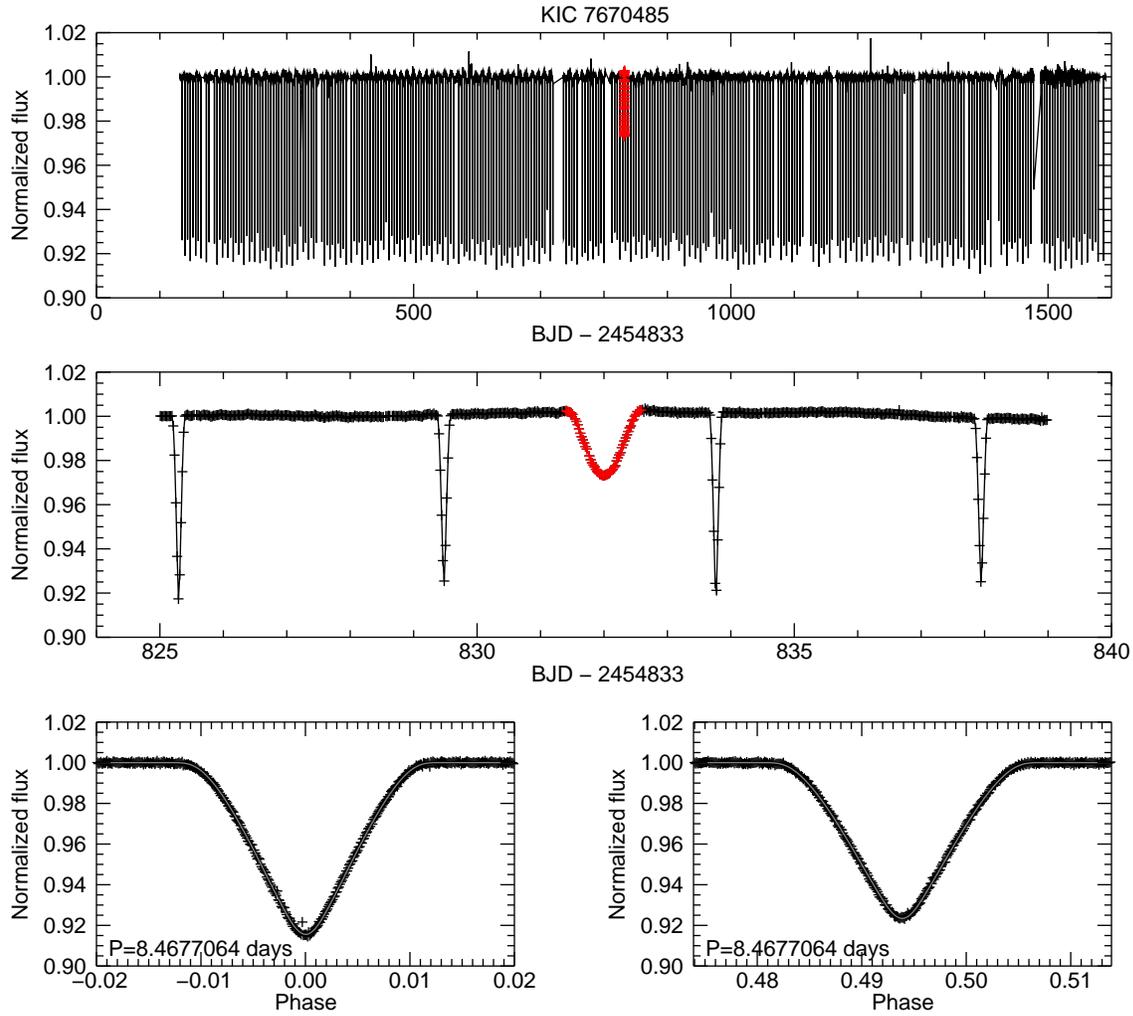}
\caption{Upper: The whole light curves of KIC 7670485 with one extra eclipse in red plus. Lower: The expanded region around the extra eclipse.} \label{KIC007670485.eps}
\end{figure}

Not many papers were published on the object KIC 7670485 presently. Apart from the Kepler Eclipsing Binary Stars catalog of  Pr{\v s}a et al. (2011), Slawson et al. (2011), Conroy et al. (2015) and Kirk et al. (2016), Tenenbaum et al. (2012) detected its transit signals in the first three quarters of \textsl{Kepler} Data, and Armstrong et al. (2014) provided its temperatures from three photometric surveys data. Here a clear and credible extra eclipse is found at 832 days, see Figure \ref{KIC007670485.eps}. The significant and symmetrical contour of the eclipse strongly suggest the existence of the additional stellar body. However, only one signal is not enough to obtain the physical parameter behind. The additional stellar body could be a third one around the inner binary, or another binary system, gravitational bound or blending with the existing binary. No visible eclipse timing variations are found so far, the actual dispersion of its $O-C$ curve measured by us is less than 0.001 days. This shows that if the eclipse on 832 days is due to a revolving third stellar, its orbital period should be much greater than 1500 days.

The inner 8.4677064 days period binary was analyzed, and the results are shown in the lower panels of Figure \ref{KIC007670485.eps} and listed in Table \ref{tab:binary_solution}. The mass ratio in the solution has large uncertainty, because a large range of mass ratios can give roughly the same fitting. Taking into account the other parameter ratios are also around 1, so the solution with mass ratio around 1 is given here. The third light proportion is dominating the whole luminosity, which indicating that the extra body (or bodies) should exist, and it is much brighter than the inner binary. Based on the large third light proportion, a binary system with greater mass is suggested to be the extra bodies.

\subsection{KIC 8938628}

\begin{figure}
\includegraphics[scale=.50]{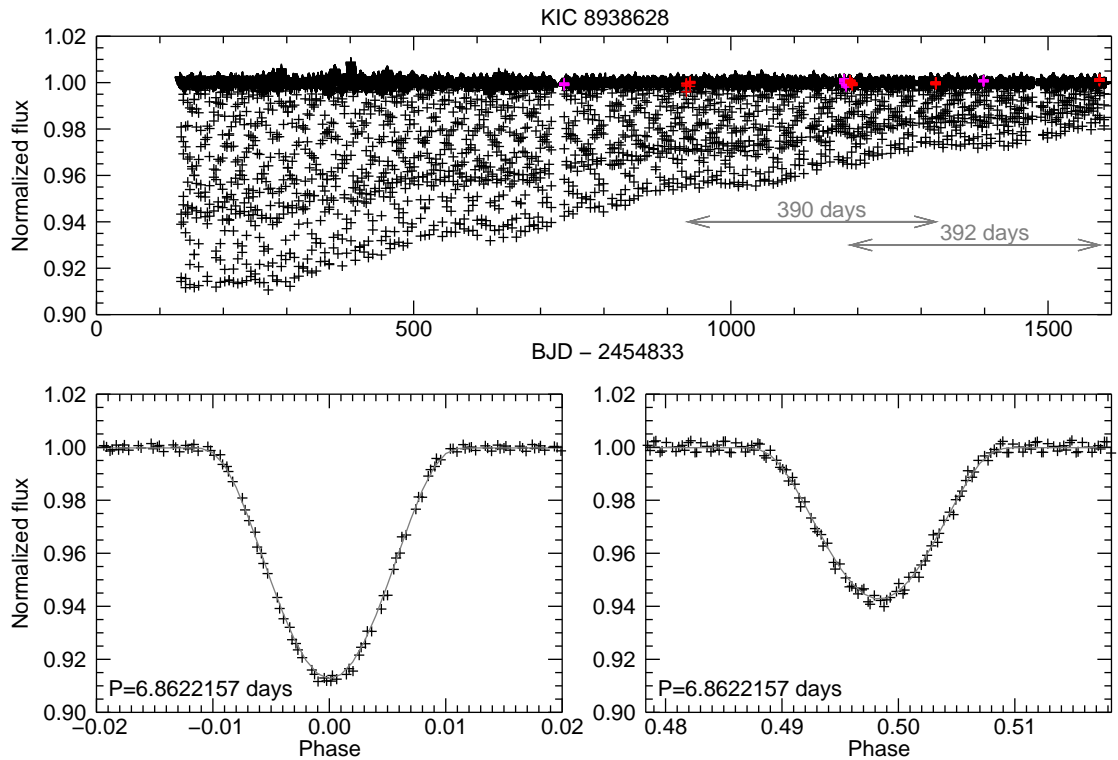}
\caption{Upper: The whole light curve of KIC 8938628, the red and magenta points are the extraneous eclipses, and the gray arrows are used to indicate the time intervals of the red points. Lower: The phase curves of the inner 6.8622157 days period binary with gray fitting lines.} \label{KIC008938628.eps}
\end{figure}

KIC 8938628 was found as an eclipsing binary system by Pr{\v s}a et al. (2011) and Slawson et al. (2011). Rappaport et al. (2013) and Borkovits et al. (2015) analyzed its monoperiodic large amplitude $O-C$ curve and concluded that a $1.12\pm0.28 M_{\odot}$ third stellar, rotating around the nearly circular orbital inner binary, in $388.5\pm0.3$ days period. The obvious variation of the eclipse depth can be seen from the upper panel of Figure \ref{KIC008938628.eps}, which indicates the varying inclination of the inner binary due to a not coplanar third body around (Borkovits et al. 2015). If the decreasing tendency continues, the binary eclipses will disappear at about $BJD \sim2456813$, or 4 June 2014, like the case of KIC010319590 which eclipse depths completely disappear in $\sim400$ days (Rappaport et al. 2013).

\begin{figure}
\plottwo{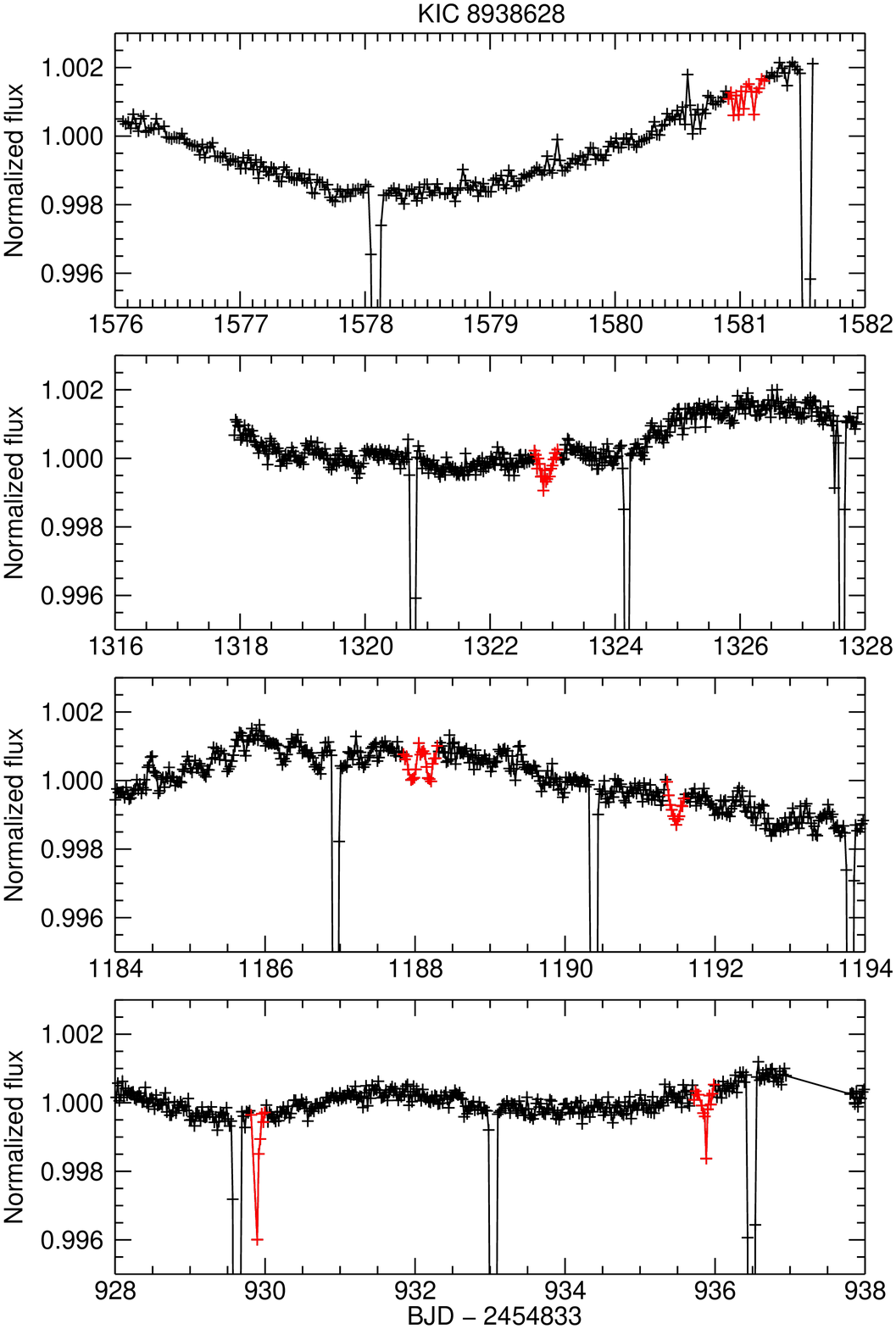}{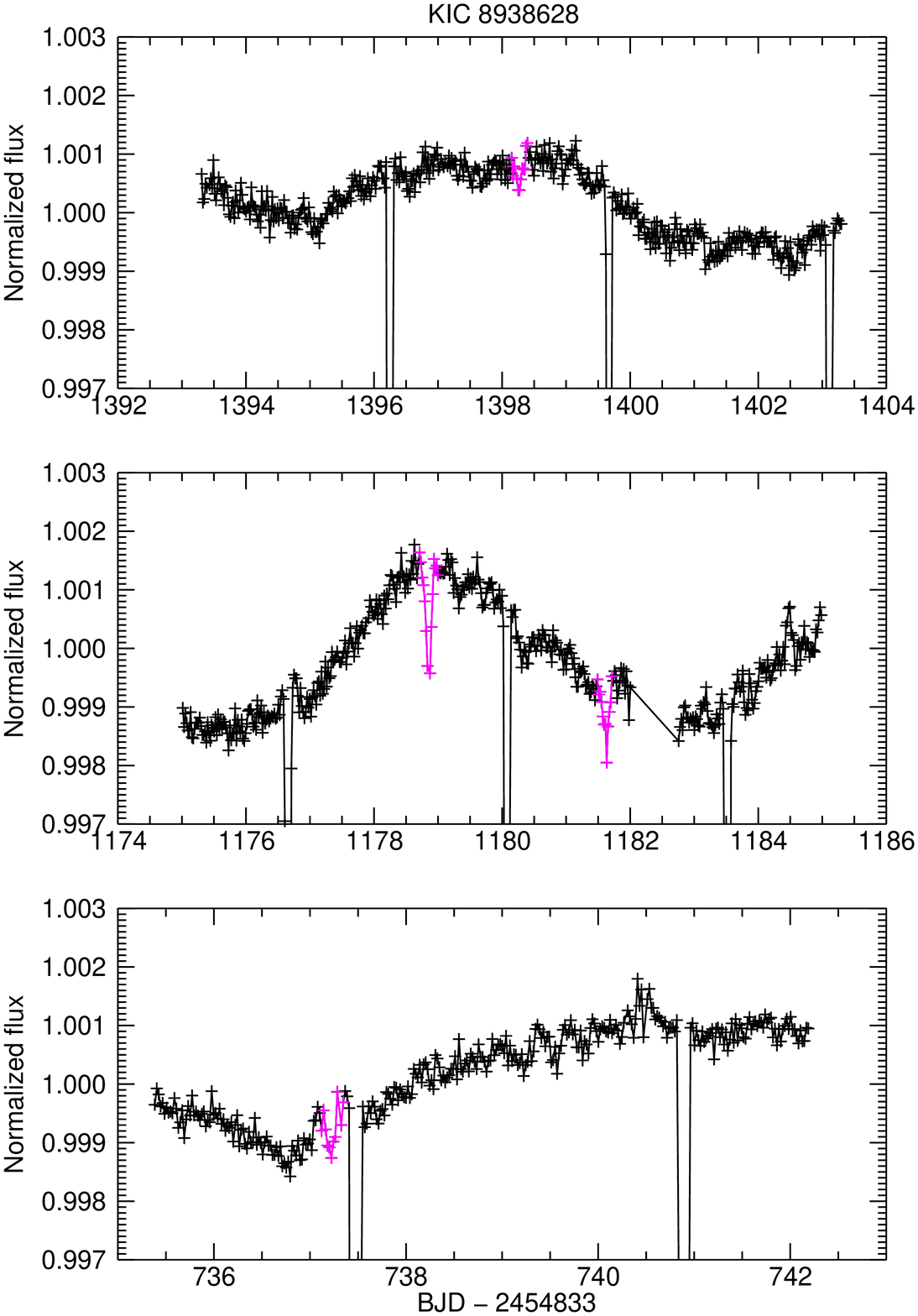}
\caption{The extraneous eclipses of KIC 8938628. Left panels: the red points are the extraneous eclipses. The time interval between the extraneous eclipses in the first and third panel from top is about 390 days, which is close to that between the second and fourth panel that is about 392 days. Right panels: the other set of extraneous eclipses in magenta, the time intervals between them are about, or times of, 220 days.} \label{KIC008938628_extra_eclipses_1_and_2.eps}
\end{figure}

Many small dips of suspected tertiary eclipses are found on its light curves, and the most notable ones are shown in Figure \ref{KIC008938628_extra_eclipses_1_and_2.eps}.\footnote{The data for this object KIC 8938628 are the version of August 1, 2014. While the differences between the different versions are found like KIC 7668648. At around 1179 days, an evident extra eclipse was found in the version of August 1, 2014, but disappeared in the version of April 22, 2015 or later.} The extraneous eclipses are only suspected but not substantial for their low signal to noise ratio. However, the period information carried by the extraneous eclipses are worthy of attentions. The extraneous eclipses in red, see the upper panel of Figure \ref{KIC008938628.eps}, imply an eccentric orbit in about 390 days period, which is very close to the $388.5\pm0.3$ days period of third body derived by the periodic analysis method (Borkovits et al. 2015). Furthermore, the extraneous eclipses in magenta also exhibit a pattern pointing a period of $\sim220$ days. It is suggested that the object KIC 8938628 is a potential multiple eclipsing system worthy follow-up observations.

The solution of photometric analysis of the inner 6.8622157 days period binary is displayed in Figure \ref{KIC008938628.eps} and Table \ref{tab:binary_solution}. The extra luminosity only occupies a proportion of 1/3, which indicates that the extra bodies may consist of several small components so their total luminosity is low, and this is also conform to the tiny extraneous eclipses presented above. The object KIC 8938628 is suggested to a hierarchical quadruple system candidates.

\section{summary and discussion}

In this paper four \textsl{Kepler} binaries with 57 extraneous eclipses, indicating the additional bodies around the binaries, are presented. All the objects binary light curves are analysed using the binary model Wilson-Devinney programme with good fittings. Three of the four multiple system candidates are (at least) quadruple systems, and one of them is triple system.

\object{KIC 7622486} is a double binary system with two strikingly sets of eclipses with period of 2.2799960 days and 40.246503 days period. The two binary systems are thought to be gravitational bound to each other based on the existence of a small eccentricity 0.0035(0.0022) in 2.2799960 days period orbit. This system provide a very good window to study the formation of multiple system as well as binary system. If the two binaries share with the same age, they should form from the same collapsing cloud. Otherwise if the two binaries have different ages, the \object{KIC 7622486} is a strong support to the theory of gravitational capture with a experience of unstable three body stage and one of the component will be ejected from the remaining stable binary system.

\object{KIC 7668648} and \object{KIC 8938628} are hierarchical quadruple candidates with more than one set of extraneous eclipses. Their tripility are found by Rappaport et al. (2013) and Borkovits et al. (2015) using the $O-C$ method. The first two sets of extraneous eclipses on KIC 7668648 and the first set of extraneous eclipses on KIC 8938628 firmly confirm the their tripility with independent proofs. Moreover, the last set of extraneous eclipses on KIC 7668648 and KIC 8938628 indicates additional body (or bodies) within the multiple systems, which make the two objects more interesting and more challenging.

For KIC 8938628, the last set of extraneous eclipses, shown in the right panels of Figure \ref{KIC008938628_extra_eclipses_1_and_2.eps}, may contain a period of about 220 days. If it is true, because the 220 days period of the fourth body does not differ greatly with the $388.5\pm0.3$ days period of the third body, the system can not be stable unless the mass of the fourth body is marginal compared to other components. The fourth body may be a planet within a triple stellar system. If the 220 days period is not true, it needs more than one additional bodies to generate the last set of extraneous eclipses which make this object a quinary candidates. It should be noted that the 220 days period can not be from a binary system, because there are two eclipses in one group (see the middle one of the right panels of Figure \ref{KIC008938628_extra_eclipses_1_and_2.eps}) which can be only from a triple eclipse structure.

Only one but very credible extraneous eclipse found on KIC 7670485's binary light curves, and this make it a triple system candidate. No further meaningful information can be obtained from only one extraneous eclipse.

The high proportion of systems carrying more than one additional companions (i.e. the quadruple or higher order system, here in our four objects the proportion is $3/4$) is a hint to us that the multiple systems may have the tendency of forming with more than three components, and also reminds us that the triple systems has been found currently are likely to have more companions. This is similar to the case of exoplanet candidates found in \textsl{Kepler} data that approximately 40\% of the candidates belong to multi-planet systems (Rowe et al. 2014).

\begin{acknowledgements}
This paper includes data collected by the \textsl{Kepler} mission. Funding for the \textsl{Kepler} mission is provided by the NASA Science Mission Directorate. This work is partly supported by the West Light Foundation of The Chinese Academy of Sciences, Yunnan Natural Science Foundation (Y5XB071001), Chinese Natural Science Foundation (No. 11133007 and No. 11325315), the Key Research Program of the Chinese Academy of Sciences (grant No. KGZD-EW-603), the Science Foundation of Yunnan Province (No. 2012HC011), and by the Strategic Priority Research Program ``The Emergence of Cosmological Structures'' of the Chinese Academy of Sciences (No.XDB09010202).
\end{acknowledgements}

\label{lastpage}

\end{document}